\documentclass[11pt,a4paper]{article}

\usepackage{graphicx}
\title{Jet production and measurements of $\alpha_S$ at HERA\footnote{presented at Int. Europhysics Conference on High Energy
  Physics, July 21st - 27th 2005, Lisboa, Portugal}
}

\author{
  Arnd Specka\footnote{behalf of the H1 and ZEUS collaborations}\\
  Laboratoire Leprince-Ringuet, Ecole Polytechnique, France
}

\newcommand{\as}{\ensuremath{\alpha_S\,}}
\newcommand{\asn}[1]{\ensuremath{\alpha_S^{#1}\,}}
\newcommand{\asmz}{\ensuremath{\mathrm{\alpha_S(m_Z)\,}}}

\newcommand{\ETB}{\ensuremath{\mathrm{E_{T,B}}}}
\newcommand{\etaB}{\ensuremath{\mathrm{\eta_{T,B}}}}
\newcommand{\uR}{\ensuremath{\mathrm{\mu_R}}}
\newcommand{\uF}{\ensuremath{\mathrm{\mu_F}}}
\newcommand{\QQ}{\ensuremath{\mathrm{Q^2}}}
\newcommand{\GeV}{\ensuremath{\textrm{GeV}}}
\newcommand{\GeVV}{\ensuremath{\textrm{GeV}^2}}
\newcommand{\AsymErr}[2]{
  \mbox{\footnotesize$\begin{array}{l}\mathrm{#1}\\\mathrm{#2}\end{array}$}
}

\begin{document}
\maketitle

\begin{abstract} 
  The inclusive jet, dijet and trijet production cross-sections in deep-inelastic
  scattering (DIS) at $\sqrt{s}=320$~\GeV\ have been measured at the electron-proton collider HERA
  by the H1 and ZEUS collaborations using data taken in 1998--2000. The jet cross-sections have been
  measured differentially in four-momentum transfer squared \QQ\ and jet transverse energy in the
  Breit-frame \ETB\ with a typical precision of 5--10\% limited by systematic uncertainties such as
  hadronization corrections and hadronic energy scale.  All jet observables are well described by
  perturbative QCD (pQCD) predictions at next-to-leading order (NLO) within the estimated accuracy
  of these calculations which is limited by the absence higher orders, and which is in general
  inferior to the experimental precision. The values for the coupling constant \as\ of the strong
  interaction as determined by fits of pQCD predictions to the inclusive jet cross-section and the
  trijet-to-dijet production ratio $R_{3/2}$ are consistent for both experiments and both
  observables, and also are in excellent agreement with the world average.  Combining the individual
  measurements, a common value of $\as=0.1186 \pm 0.0011 \mathrm{(experiment)} \pm 0.0050
  \mathrm{(theory)}$ is obtained.  
  \end{abstract}

\section{Introduction}
One of the many ways to determine \as\ is through the measurement of jet production in DIS since the
production cross section in leading order of $n$ jets (not counting the proton remnant) is
proportional to \asn{n-1}. At HERA this measurement is particularly interesting since the hadronic
final state is opened up in pseudo-rapidity. The measurements of jet production cross-section
presented here have many points in common which facilitates comparison of the extracted values of
\as. In particular, jets are measured throughout using the longitudinally invariant $k_T$ algorithm
in the Breit-frame of reference, where in the naive quark parton model the struck quark and the
virtual boson collide head-on. Hence, substantial transverse energy \ETB\ in the Breit frame stems
necessarily from QCD radiation and provides a natural energy scale for the determination of \as.

\section{Measurements of inclusive and multi-jet cross-sections in DIS}

\begin{figure}[p]
  \centering
  \includegraphics[bb=0 0 500 350,viewport=0 0 500 350, clip=true,width=0.95\textwidth]{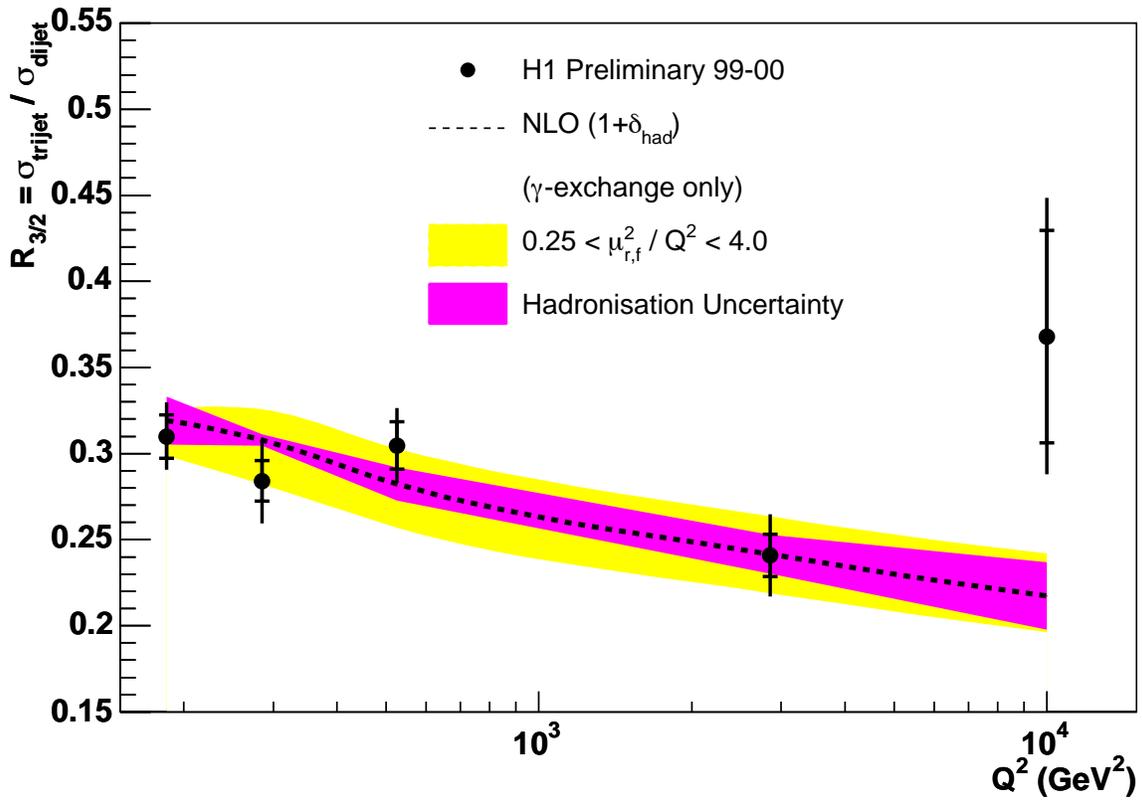}
  \caption{Trijet/dijet production ratio $R_{3/2}$ (H1)}
  \label{Fig:HMulJetR32}
\end{figure}

In the H1 (ZEUS) inclusive-jet analysis, all jets in DIS events with $150<\QQ<5000$~\GeVV
($\QQ>125$~\GeVV) are counted if their transverse energy \ETB\ exceeds 7~\GeV (8~\GeV).
Furthermore, the pseudo-rapidity of each jet in the laboratory frame (Breit frame) is required to
lie between -1.0 and 2.5 (-2.0 and 1.5) in the H1 (ZEUS) analysis.  The inclusive jet cross-section
is measured single- and double-dif\-fer\-en\-ti\-al\-ly in \QQ\ and in \ETB, and also differentially
in the pseudo-rapidity \etaB\ of the jet in the Breit frame (ZEUS
analysis)~\cite{Ref:HInc,Ref:ZInc}.  For the measurement of dijet and trijet cross-sections a lower
minimal transverse energy of $\ETB>5$~\GeV\ is required in both analyses, and jets in the same
pseudo-rapidity interval as for the H1 inclusive jet analysis are
considered. In order to ensure infrared-safeness of the QCD predictions for
multi-jet cross-sections, the invariant mass of the 2 or 3 jets is required to be above 25~\GeV. The
H1 and ZEUS analyses differ in the kinematic range of the DIS events: $150<\QQ<15000$~\GeVV\ for H1
and $10<\QQ<5000$~\GeVV\ for ZEUS. The multi-jet cross-sections are measured differentially in \QQ\ 
(H1) and \QQ, \ETB, and \etaB (ZEUS), and the ratio $R_{3/2}$ of the trijet and the dijet
cross-sections is measured as a function of \QQ\ (fig.~\ref{Fig:HMulJetR32})~\cite{Ref:HMul,Ref:ZMul}.

In all analyses, the data are corrected bin-by-bin for detector effects and QED radiation using
fully simulated Monte Carlo events, using leading-order (LO) event generators. The dependence of the
detector correction on the different models for parton showering --- leading-log parton showers (as
implemented in RAPGAP and LEPTO) and the colour-dipole model (as implemented in ARIADNE) --- results
in a systematic uncertainty of e.g. $\pm 7\%$ for the inclusive jet cross-section (ZEUS).  A
systematic error of comparable size stems from the uncertainty of the absolute energy
scale of the hadronic calorimeters. It is evaluated by varying the calorimeter energy scale by $\pm
2\%$ for all events and fully resconstructing these (H1) or simply by varying the transverse energy
of the jets' \ETB\ by $\pm$1--3\% depending on \ETB\ (ZEUS). Typically, this results in an
uncertainty of $5\%$ on the inclusive-jet cross-section as measured by ZEUS and of 1\% (4\%) on the
dijet (trijet) cross-section as measured by H1.  Other experimental uncertainties, such as
systematic errors on the measurement of luminosity or on the reconstruction of the scattered
electron, typically contribute to the overall experimental error on a level of 1--2 per-cent.

\section{Comparison with perturbative QCD at next-to-leading order}
The measured jet cross-sections are compared to NLO pQCD predictions calculated using the programs
DISENT (${ \cal O}(\as^2)$) and NLOJET++ (${\cal O}(\as^3)$). However, these codes do not include
parton-showering and fragmentation, and deliver cross-section values for parton final states only.
In order to compare these calculations to measured data, a hadronization correction factor is
applied to the pQCD predictions as the ratio of the cross-sections for hadron and parton final
states as calculated by the above mentioned event generators. This procedure relies on the
assumption that the hadronisation correction is not significantly affected by the
difference between LO and NLO parton final states. The theoretical error on this hadronization
correction is estimated by comparing the correction coefficients obtained with the two different
models for parton showering.

\begin{figure}[p]
  \centering
  \includegraphics[bb=42 0 462 568,clip=true, width=0.8\textwidth]{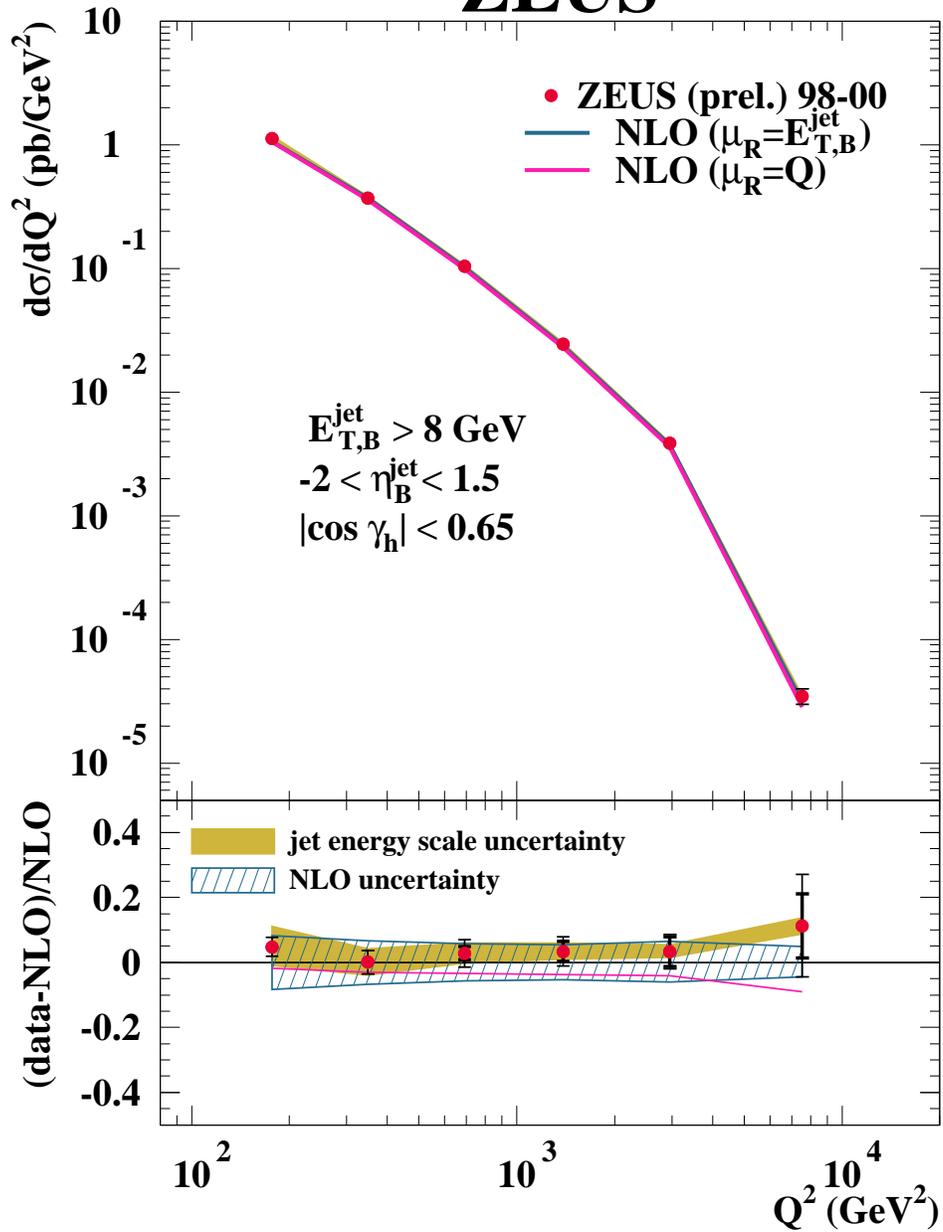}
  \caption{Differential inclusive Jet Cross-Section in \QQ}
  \label{Fig:ZIncXsctQ2}
\end{figure}

Perturbative QCD calculations at fixed order unavoidably exhibit an explicit dependence on the
renormalisation scale \uR reflecting the absence of higher orders in the perturbative
expansion. It has been shown in these analyses that the QCD predictions for the discussed jet
observables are not very sensitive to the particular choice for \uR\ and the factorization scale
\uF\ among the possible scales of each event:\QQ, \ETB, or an arithmetic combination of both, but
rather to the absolute value of \uR (cf. fig.~\ref{Fig:ZIncXsctQ2}). In order to assess the
theoretical uncertainty on the jet cross-sections and \as\ due to missing higher orders, the
renormalization scale \uR\ is varied. It has to be pointed out that the particular choice of varying
\uR\ up and down factor of 2 is a mere convention - that is followed in the presented analyses - but
has no theoretical motivation.
For the inclusive jet cross-section (e.g. ZEUS), the uncertainty resulting from the renormalization
scale dependence ($\pm5\%$) dominates over the other contributions, e.g parametrization of the
proton PDF ($\pm3\%$), to the overall error of the pQCD prediction. This is also true for the
multi-jet cross-sections (e.g. H1): predicted cross-section for dijets (trijets) vary by as much as
$\pm 3\% $ ($\pm 10\%$) with \uR\ whereas the uncertainty from the hadronization correction is
$\pm 1\%$ ($\pm 4\%$).  The lower part of figure~\ref{Fig:ZIncXsctQ2} demonstrates the good
agreement of the NLO pQCD prediction with the measured inclusive jet cross-section (ZEUS) as it
drops over five orders of magnitude on a wide range in \QQ. Theory and experiment are compatible
within experimental error which are generally smaller than the estimated theoretical uncertainty.
The good agreement holds for the multi-jet rates and their ratio as well, as can be seen on
fig.~\ref{Fig:HMulJetR32} except at very high \QQ\ where electro-weak effects are
not accounted for in the NLO pQCD programs.
 
\section{Determination of the strong coupling constant \as}
\begin{figure}[p]
  \centering
  \includegraphics[width=0.95\textwidth]{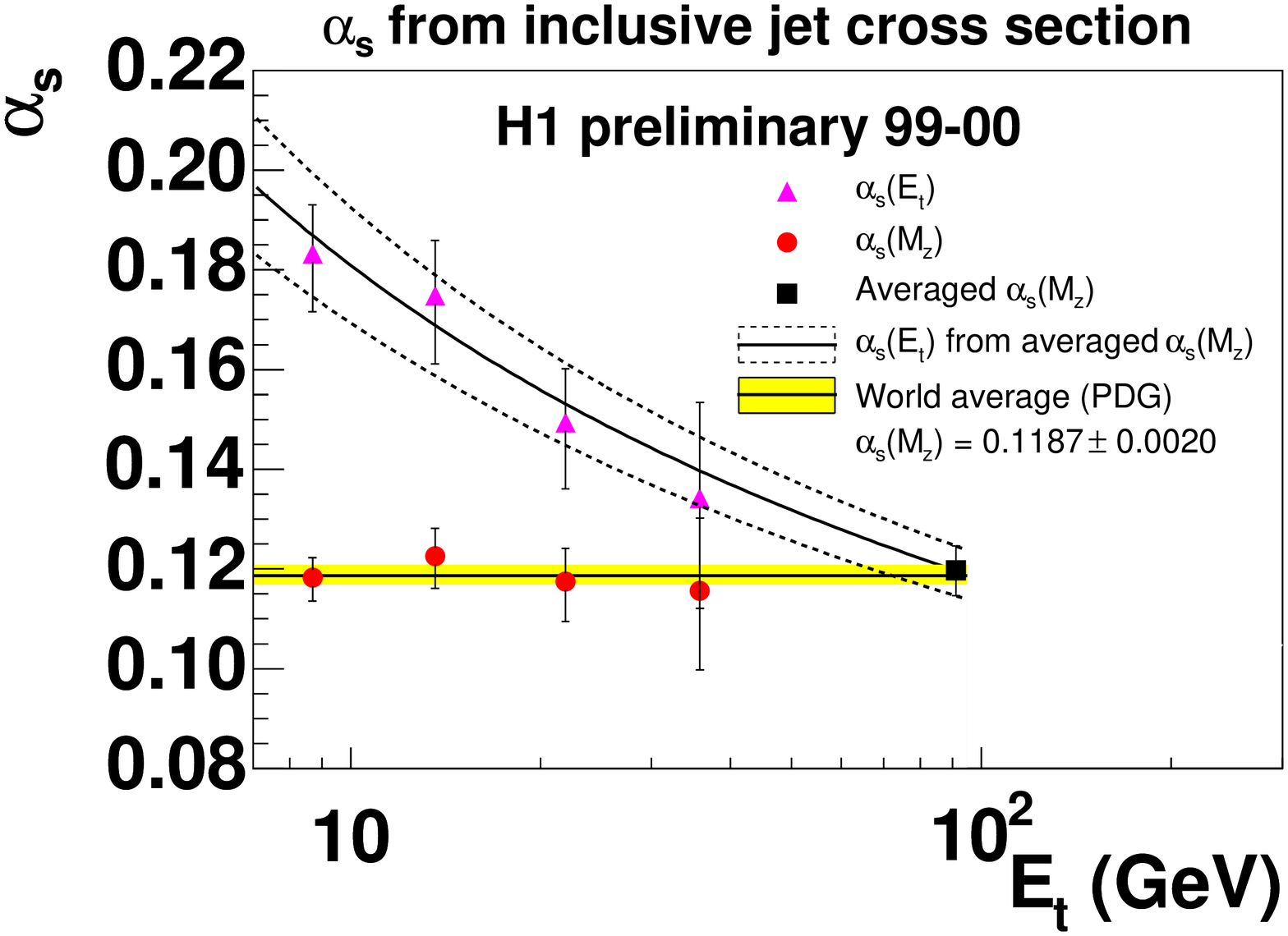}
  \caption{Running $\as(\ETB)$ from inclusive jet $\sigma$}
  \label{Fig:HIncJetAs}
\end{figure}

The strong coupling constant is determined by fitting the pQCD predictions as a function of \as to
the measured values of the jet observables, i.e. the cross-section value in a particular bin in \QQ\ 
or \ETB\ or the ratio $R_{3/2}$. In practice, the pQCD prediction for each observable is parametrised
as a second order polynomial in \as\ (with vanishing constant term).  Figure~\ref{Fig:HIncJetAs}
shows the running $\as(\ETB)$ as extracted from the differential inclusive jet cross-section in
\ETB\ (H1), and the corresponding value evolved to the Z$^0$-boson mass.  By simultaneously fitting the
observables over the range where they are well described by pQCD, and also taking
partial correllations of systematic experimental and theory errors into account, an \asmz-measurement
is extracted for each of the four analyses: 
\begin{center}
\renewcommand{\arraystretch}{0.90}
\begin{tabular}{@{\quad}llc@{\quad}r@{\quad}r@{\quad}r@{\quad}r}
  &&Ref.&\asmz&   \multicolumn{2}{c}{\footnotesize(exp. error: stat.,syst.)} & {\footnotesize(theory error)} \\\hline
   Incl. Jets        &  H1 & \cite{Ref:HInc} & 
   0.1197&
   \multicolumn{2}{c}{$\pm 0.0016$ }&
   \AsymErr{+0.0046}{-0.0048}\\
   
                     & ZEUS & \cite{Ref:ZInc} &
  0.1196&
  $\pm 0.0011$  &  
  \AsymErr{+0.0019}{-0.0025}&
  \AsymErr{+0.0029}{-0.0017}\\

  $R_{3/2}$          & H1 & \cite{Ref:HMul} & 
  0.1175&
  $\pm 0.0017$&  
  $\pm 0.0050$&
  \AsymErr{+0.0054}{-0.0068}\\

                     & ZEUS & \cite{Ref:ZMul} & 
  0.1179& 
  $\pm 0.0013$& 
  \AsymErr{+0.0028}{-0.0046}&
  \AsymErr{+0.0064}{-0.0046}\\

  \multicolumn{2}{l}{ HERA combined} &\cite{Ref:Glas} &                 
  { 0.1186} & 
  \multicolumn{2}{c}{$\pm 0.0011$}  & 
  {$ \pm 0.0050$}
\end{tabular}
\end{center}

For the same observable, both experiments obtain almost identical values of
\asmz\, with slightly lower values and higher experimental error for a
measurement using $R_{3/2}$.  The \asmz\ values obtained from different
observables, agree well with each other, with an average of previous HERA
measurements from inclusive DIS and jet data~\cite{Ref:Glas} (in
which~\cite{Ref:ZMul} is used), and also with a world average of
$0.1182\pm0.0027$ as determined by~\cite{Ref:Beth} at a level of experimental
precision competitive with the most precise measurements for jets.

It should be noted that the error from the pQCD calculation is roughly three
times higher than the experimental uncertainty for the most precise of these
four measurements~\cite{Ref:HInc}. Although the latter could probably be reduced
by increased statistics and better understanding of systematical errors, future
progress on the overall precision for \as\ hinges on an improved estimation of
contributions beyond NLO to the pQCD predictions.

\begin{thebibliography}{99}
\bibitem{Ref:HInc} H1~Collaboration, contributed paper No. 629 at this conference\\[-5mm]
\bibitem{Ref:ZInc} ZEUS~Collaboration, contributed paper No. 375 at this conference\\[-5mm]
\bibitem{Ref:HMul} H1~Collaboration, contributed paper No. 625 at this conference\\[-5mm]
\bibitem{Ref:ZMul} 
S.~Chekanov et al., 
\emph{European Physical Journal} {\bf C44} (2005) 183-193 [{\tt hep-ex/0502007}]\\[-5mm]
\bibitem{Ref:Glas} 
C.~Glasman, 
 \emph{XIII Int. Workshop on Deep Inelastic Scattering, Madison} (2005) [{\tt hep-ex/0506035}]\\[-5mm]
\bibitem{Ref:Beth} 
S.~Bethke, \emph{Nucl.Phys.Proc.Suppl. {\bf 135}} (2004) 345-352 [{\tt hep-ex/0407021}]\\[-5mm]
\end{thebibliography}
\end{document}